\documentclass[5p,times]{elsarticle}

\usepackage{amssymb}
\usepackage{amsmath}
\usepackage{tabularx}
\usepackage{hyperref}
\usepackage{algorithm}
\usepackage{algpseudocode}
\usepackage{xcolor}
\newcommand{\correction}[1]{#1}

\journal{Future Generation Computer Systems}

\begin{document}

\begin{frontmatter}



\title{Performance optimization of BLAS algorithms with band matrices for RISC-V processors}



\author{Anna Pirova}
\author{Anastasia Vodeneeva}
\author{Konstantin Kovalev}
\author{Alexander Ustinov}
\author{Evgeny Kozinov}
\author{Alexey Liniov}
\author{Valentin Volokitin}
\author{Iosif Meyerov\corref{cor1}}

\cortext[cor1]{Corresponding author.}
\ead{meerov@vmk.unn.ru}

\affiliation{organization={Department of HPC and System Programming, Lobachevsky State University of Nizhny Novgorod},
            addressline={23, Prospekt Gagarina}, 
            city={Nizhny Novgorod},
            postcode={603022}, 
            country={Russia}}

\begin{abstract}
The rapid development of RISC-V instruction set architecture presents new opportunities and challenges for software developers. Is it sufficient to simply recompile high-performance software optimized for x86-64 onto RISC-V CPUs? Are current compilers capable of effectively optimizing C and C++ codes or is it necessary to use intrinsics or assembler? Can we analyze and improve performance without well-developed profiling tools? Do standard optimization techniques work? Are there specific RISC-V features that need to be considered? These and other questions require careful consideration. In this paper, we present our experience optimizing four BLAS algorithms for band matrix operations on RISC-V processors. We demonstrate how RISC-V-optimized implementations of OpenBLAS algorithms can be significantly accelerated through improved vectorization of computationally intensive loops. Experiments on Lichee Pi 4A and Banana Pi BPI-F3 devices using RVV 0.7.1 and RVV 1.0 vector instruction sets respectively, show speedups of 1.5x to 10x depending on the operation compared to the OpenBLAS baseline. In particular, the successful use of vector register grouping with RVV can lead to significant performance improvements.
\end{abstract}






\end{frontmatter}



\section{Introduction}
\label{sec1}

The development of the RISC-V instruction set architecture (ISA) opens up new opportunities to solve a wide range of problems focusing on high-performance computing systems. Openness, freedom from patent restrictions and unlimited possibilities for further development \cite{c1} attract researchers and engineers around the world. In fact, there are no high-performance server-class systems that are competitive in performance with top x86 architecture solutions. Despite this, the current level of development in this area allows us not only to dream about the future but also to conduct quite successful research and development. This fact is reflected in the emergence of thematic workshops on RISC-V HPC at major conferences on supercomputer technologies, such as, for example, Supercomputing Conference, ISC High Performance, Parallel Programming and Applied Mathematics, High Performance, Edge And Cloud Computing. Apparently, in the coming years we could expect the emergence of multicore vector out-of-order RISC-V processors suitable for building HPC-class systems. The use of such systems requires not only the availability of appropriate equipment, but also the development of the entire ecosystem. The current state of affairs in the readiness of the RISC-V ecosystem area for HPC is described in \cite{c2}. The paper \cite{c3} provides an overview of open source RISC-V architecture processor cores. The work \cite{c4} examines existing and future RISC-V ISA extensions and demonstrates exciting prospects for the creation and use of RISC-V devices for various application areas (IoT, AI, HPC, Communication, Security, Computer Graphics, etc.). 

Further progress is impossible without the development of system software. A systematic review of the achievements in the area of RISC-V system software is presented in \cite{c5}. The work \cite{c6} describes the Vitamin-V project, which is being developed with the support of the Horizon Europe program. The goal of the project is to create a comprehensive open-source software stack for RISC-V that can be used in cloud services \cite{c6}. Great opportunities for the RISC-V community lie at the intersection of hardware and software, in the area of software-hardware co-design. The papers \cite{c7} and \cite{c8} discuss the exceptional importance of predictive modeling of processor architectures and early feedback on proposed technical solutions in the creation of future supercomputers. The works \cite{c9} and \cite{c10} clearly demonstrate the possibilities of co-design using two algorithms: multiplication of sparse matrices by dense vectors and convolution. The future success of this approach is undoubted. A vast body of work has been devoted to analyzing performance and optimizing programs for current and future RISC-V architecture processors. In \cite{c11} and \cite{c12}, a detailed study of the 64-core SG2042 processor performance was carried out -- the first processor with dozens of RISC-V cores claiming to belong to the HPC domain. The authors demonstrated that the processor, although significantly inferior to its x86 counterparts, demonstrated acceptable performance for compute-bound applications while in memory-bound scenarios it required improved memory handling mechanisms. The papers \cite{c13,c14,c15} discussed the results of testing RISC-V devices using tests typical of the HPC area. The paper \cite{c16} provided an overview of the state-of-the-art in vectorization on RISC-V and analyzed the capabilities of Allwinner D1 processors. The papers \cite{c17} and \cite{c18} showed the advantages of using long SIMD vectors on promising RISC-V processors. The paper \cite{c19} studied the capabilities of compilers for automatic vectorization. A series of papers are devoted to testing the performance and optimization of specific scientific software in astrophysics \cite{c20}, \correction{computational fluid dynamics simulations \cite{c71},} machine learning \cite{c21}, computer vision \cite{c22}, graph algorithms \cite{c23}, optimization of Fast Fourier Transform \correction{\cite{c24,c70}}  and Merge sort \cite{c25} algorithms.

In general, the studies performed allow us to draw the following conclusions. \textit{Firstly}, there is a need for \correction{a wide range of performance counters and well-developed performance analysis tools \cite{c72}}. \textit{Secondly}, the experience of porting and optimizing both large problem-oriented software frameworks and individual computational kernels shows that the codes are quite easily adapted for RISC-V, and known approaches to speed up computations work as expected. \textit{Thirdly}, there is an obvious need for high-performance mathematical libraries that are optimized deeply for RISC-V processors. In this paper, we focus on optimizing the performance of four algorithms in the OpenBLAS library, one of the most common open implementations for BLAS specifications. These algorithms work with band matrices. \correction{Such algorithms are in demand for solving many problems using numerical analysis methods in scientific and engineering applications. In particular, band matrices arise in grid methods for solving the Navier -- Stokes equation in hydrodynamics, in solving the Laplace equation and the Poisson equation in problems of continuum mechanics, in solving the wave equation in problems of electrodynamics, in finite element analysis in problems of theoretical mechanics, and in many other applications.} They were chosen as targets for optimization based on large-scale performance testing of OpenBLAS on x86 processors. This showed a significant lag in these functions compared to implementations from Intel OneAPI Math Kernel Library (MKL). Therefore, the possibility of further optimization is focused primarily on the RISC-V platform, which involves improving vectorization using RISC-V specifics. The implementation uses RVV 0.7.1 and RVV 1.0 intrinsic functions, and integration with OpenBLAS is planned. The code is publicly available. 

The paper is structured as follows. Section \ref{sec2} provides an overview of publications devoted to optimizing BLAS algorithms for various hardware platforms. Section \ref{sec3} describes the main algorithms and optimizations that we implemented. Section \ref{sec4} presents the results of the computational experiments. Section \ref{sec5} concludes the paper.

\section{Related Work}
\label{sec2}

The BLAS standard was proposed in 1979 \cite{c27} and was updated until the early 2000s
\cite{c28,c29}. Dense linear algebra algorithms are the computational kernels of many libraries used in scientific and engineering applications, so a large number of works are devoted to the issues of high-performance implementation of BLAS for different architectures. There is a lot of attention in the community to optimizing the general matrix multiplication (GEMM) algorithm. This algorithm is one of the main computational kernels in many Computational Science problems. Firstly, other BLAS level 3 functions can be implemented using this algorithm \cite{c48}. Secondly, GEMM, along with other BLAS algorithms, plays an important role in other matrix algorithms in the LAPACK computational linear algebra library \cite{c65}. The most common approach to developing a high-performance GEMM implementation is described in \cite{c46}. The main idea is based on splitting the original matrix into cache-sized blocks and allocating CPU-specific low-level microkernels implemented with intrinsics or assembler. This approach was implemented in GotoBLAS and was then used in many other libraries. Optimization of other BLAS level 3 functions in GotoBLAS is described in \cite{c47}. The specifics of BLAS implementation using isolated microkernels in the BLIS library are described in \cite{c62}. For many years, high-performance implementations of GEMM and other BLAS functions have been developed by Intel in the Intel Math Kernel Library (MKL) \cite{c66}. High-performance GEMM implementations for multicore processors are discussed in \cite{c49}. With the advent of the first GPUs, approaches to GEMM optimization for graphics processors began to actively develop \cite{c51,c52}, where a fairly high degree of utilization of computing resources was also achieved. 

Optimization of vector and matrix operations for RISC-V processors is often discussed in the community in terms of improving the performance of neural network inference. In \cite{c54}, the adaptation of the integer GEMM BLIS algorithm for these purposes is proposed, and its performance on the Greenwave GAP8 processor with hardware support of dot product kernel is studied. In \cite{c55}, vectorized GEMM microkernels for RISC-V processors are generated automatically, and their performance compared to that of OpenBLAS implementations on C906 and C910 processors is evaluated. Few works have been done on optimizing algorithms for band matrices. The optimization of the band matrix-matrix multiplication algorithm has been considered in \cite{c30,c31,c32,c33,c34,c35,c36}. In early works \cite{c30,c31,c32} a matrix storage scheme based on diagonals is proposed, which differs from those used in BLAS and LAPACK, and adapted the band matrix multiplication algorithm for it. An important step towards vector computations for algorithms working with band matrices on the CRAY X-MP computer was made in \cite{c32}. In \cite{c33}, the functions of multiplying a symmetric band matrix by a vector (SBMV) and by a general matrix (SBMM) were considered. The authors proposed to split the original band matrix into dense blocks and reduce the computations to a sequence of operations with dense matrices (symmetric, general and triangular matrix-vector product for SBMV and symmetric, general, and triangular matrix-matrix product for SBMM). This approach is not always effective in the case under consideration, since we optimize the operation of OpenBLAS for matrices with a narrow bandwidth \correction{\footnote{\correction{In linear algebra, for a matrix $A=(a_{ij})$ the bandwidth is the maximum distance between two non-zero elements in a row: $\max_{a_{ij}\not=0}|i-j|$; for a dense matrix, it is equal to the number of non-zero diagonals of the matrix.}}}. In \cite{c34}, this approach was adapted for use on GPUs, and the symmetric band matrix storage scheme was modified to reduce the number of function calls from BLAS level 2. In \cite{c35,c36}, a compact diagonal band matrix storage scheme and the corresponding band matrix-matrix multiplication algorithm for CPUs \cite{c35} and GPUs \cite{c36} are proposed. In \cite{c58}, a multithreaded algorithm for band matrix-vector multiplication is presented, which guarantees the same computational accuracy for multiple runs of the application.

With the advent of the new RISC-V architecture, the issue of the basic algorithms implementations efficiency, taking into account the requirements and limitations of the BLAS specifications, is once again becoming relevant. These include algorithms for working with band matrices. To study this, it was necessary to choose one of the BLAS implementations optimized for RISC-V processors with RVV 0.7.1 or RVV 1.0 extensions. Currently, there are many open source and commercial implementations available with low-level optimization for various architectures. Some libraries are optimized for specific types of processors, while others have implementations for a wide range of architectures. Thus, the free software package Intel OneAPI Math Kernel Library \cite{c37} is widely used for Intel processors, the AOCL-BLAS package \cite{c39} optimized for AMD processors and IBM ESSL libraries optimized for IBM processors are also popular. Open packages such as ATLAS \cite{c38}, ARMAS \cite{c43}, BLIS \cite{c62}, Eigen \cite{c41}, and OpenBLAS \cite{c40} are optimized for many different platforms and are widely used in scientific research. One of the newest libraries, BLASFEO \cite{c44}, is optimized for embedded systems and provides a high-performance BLAS implementation for cache-fitting matrices. The Netlib implementation is often used as the baseline reference implementation for BLAS on x86 and some other platforms \cite{c59}. For GPUs, libraries such as cuBLAS, clBLAST, ViennaCL are commonly used. The BLIS, OpenBLAS, and Eigen libraries have low-level computational kernels for a wide range of processors, including ARM, IBM Power, MIPS (OpenBLAS, Eigen), but vectorization for RISC-V processors is only partially supported. A comparison of various BLAS implementations on ARM and RISC-V processors as of 2020 is given in \cite{c26}. The BLIS library implements vectorization for SiFive x280 processors (August 2024). The latest release of the Eigen library does not support RISC-V, but a merge request with support of RVV 1.0 instructions (September 2024) is added. Compared to other libraries, OpenBLAS has the most complete vectorization for RISC-V processors and low-level computational kernels with RVV 0.7.1 and RVV 1.0 support for several types of processors. Therefore, we chose it as a baseline implementation, studied the performance of the developed algorithms and identified functions that had potential for further improvement.

\section{Algorithms and their implementation for RISC-V CPUs}
\label{sec3}

\subsection{Methodology of selecting algorithms for optimization}

As the review of BLAS implementations has shown, the OpenBLAS library is widely used in computations and has the most complete implementation for RISC-V processors. \correction{It should be noted that C++ compilers are not yet able to automatically vectorize loops for the RVV 0.7.1 vector instruction set. However, modern compilers are already capable of automatically vectorizing loops for RVV 1.0, and using autovectorization is a basic recommendation for many applications. When developing high-performance libraries, it may be justified to use intrinsics that allow more precise control over the code and adapt the implementation to the specifics of the target architecture.} The latter consideration is also true for vector extensions available in commodity CPUs. In this regard, developers of high-performance mathematical libraries often resort to a low-level implementation of the main computational kernels using intrinsics or assembler. The OpenBLAS library also contains separate implementations using RVV 0.7.1 and RVV 1.0 intrinsics. However, many codes have potential for additional optimization \cite{c63}, and OpenBLAS is not an exception.

To investigate this issue, it was necessary to determine which of the OpenBLAS algorithms could be improved. Currently, there is no generally accepted reference high-performance BLAS implementation for RISC-V processors. Therefore, to identify algorithms with potential for optimization, we selected Intel Xeon processors with x86-64 architecture as a testing environment and compared the performance of OpenBLAS algorithms and their Intel MKL counterparts, which are highly optimized for these processors. Testing was performed for BLAS level 2 and level 3 functions in double and single precision in single-threaded and multi-threaded modes. Three classes of tests were performed based on the input data size: small, medium, and large. In the ``small'' class, the amount of initial data was chosen so that all the data fit into the L1 cache. In the ``medium'' class, it was chosen so that some data would not fit into the L1 cache but would still be within the L3 cache. In the ``large'' class, the initial data volume exceeded the L3 cache capacity, but the data still completely contained to RAM.

The BLAS-Tester library \cite{c61} was used to measure performance. Testing was repeated several times, and the time taken and the performance in MFlops were recorded. For each function, the lowest 20\% and highest 5\% performance scores were discarded from the results of each class. The remaining measurements were then averaged.

As a result of the testing, it was found that the implementation of four OpenBLAS functions that perform operations with band matrices lags significantly behind their counterparts from Intel MKL. These functions perform band matrix-vector multiplication for general, symmetric, and triangular matrices, and solve a system of linear algebraic equations with a triangular band matrix. The largest lag was observed for matrices with a narrow bandwidth, and these functions were selected for further optimization.

\correction{
\subsection{RISC-V-specific vectorization}
\label{newsec} }
\correction{The rational use of vectorization in computationally-intensive software has become one of the key methods for achieving the best performance in recent decades. It is not surprising that the features of vectorization of computations for RISC-V processors deserve close attention. In this section, we briefly review the main features of vectorization for RISC-V (RVV), assuming that the reader is generally familiar with the SIMD paradigm and the organization of vector extensions, for example, in the x86 architecture.}

\correction{\textit{The first} distinctive feature of RVV is the organisation of vector processing in loops, the number of iterations of which is not less than the number of elements that fit into the vector register. In x86, it was common to separate the main part of the loop and the "remainder" (remainder loop), which had to be considered separately. Optimizing compilers handled this successfully, but when programming in intrinsics, it was necessary to process the remainder loop manually, which increased the size of the code, the time for its development and subsequent modification. Vectorization for RISC-V assumes that the developer can calculate the size of the next portion of data and set the number of vector register elements that require processing (instructions vsetvli / vsetivli / vsetvl). In this situation, there is no need to process the "remainder" separately, which simplifies the code logic. However, there is a need to set the "real" length of the vector, and the developer (when programming in intrinsics) or the compiler (when using autovectorization) is responsible for eliminating unnecessary vsetvli / vsetivli / vsetvl instructions to improve performance. Current versions of compilers cope with this quite successfully.
}

\correction{\textit{The second} important feature is the ability to group vector registers and execute vector instructions on a group of registers combined into one "virtual" register. This is organized as follows. RISC-V ISA provides 32 vector registers, which can be considered individually (LMUL=1), in pairs (LMUL =2 corresponds to 16 large registers), in groups of four (LMUL =4 corresponds to 8 large registers), and in groups of eight (LMUL=8 corresponds to only 4 large registers). A positive feature of register grouping is the reduction in the number of instructions retired, which, together with the implementation features, can lead to a significant acceleration of calculations. The negative factor is related to the reduction in the number of available registers, which can be critical for many algorithms. Based on our experience in this and previous studies, the optimal LMUL value had to be determined experimentally. In the future, this looks like a challenge for compiler developers who are expected to implement cost models that can realistically estimate the appropriate LMUL value for different algorithms. Given that the vast majority of code is developed in high-level programming languages, the quality of the corresponding cost models will have a significant impact on performance. Apparently, current versions of compilers already contain the corresponding algorithms, but testing them on applications is a topic for a separate study.
}

\correction{\textit{The third} feature of RVV is the ability to use long vectors built into the architecture. As described above in the Introduction section, this area is being actively studied within the European RISC-V research and development program. Publications have shown the promise of this area of research, and we look forward to implementing this feature in hardware.}

\correction{Thus, there are vectorization features for RISC-V that need to be taken into account when developing programs. Currently available hardware mainly supports the RVV 0.7.1 and RVV 1.0 vector extensions, but not both. However, RVV 0.7.1 and RVV 1.0 codes can be easily converted back and forth, so although there is no talk of full portability in the usual sense of the term, it can be achieved with simple manipulations. From a performance perspective, the main difference is the twice as large vector register length in RVV 1.0, which creates additional potential for speedup in codes with regular memory access and sufficient arithmetic intensity.}

\correction{Table \ref{intrin_table} shows a short notation of the RVV intrinsics used in the pseudocode of algorithms in this paper.}

\begin{table*}[]
\caption{\correction{Employing RVV Intrinsics. The table shows the names of aliases for intrinsics (Alias), used further in the description of pseudocodes of the main algorithms, as well as their names in the RVV 0.7.1 and RVV 1.0 instruction sets. X stands for 64 for double precision variables, 32 for single precision variables, Y stands for 1, 2, 4 or 8 depending on LMUL.}}
\label{intrin_table}
\begin{center}
\begin{tabular}{ccc}
\hline
{ \textbf{Alias}}      & { \textbf{RVV 0.7.1}}                                                         & { \textbf{RVV 1.0}}                                                                              \\ \hline
{ GET\_VECTOR\_LENGTH} & { \begin{tabular}[c]{@{}c@{}}vsetvlmax\_eXmY,\\    vsetvl\_eXmY\end{tabular}} & { \begin{tabular}[c]{@{}c@{}}\_\_riscv\_vsetvlmax\_eXmY,\\ \_\_riscv\_vsetvl\_eXmY\end{tabular}} \\
{ LOAD}                & { vle64\_v\_fXmY}                                                             & { \_\_riscv\_vle64\_v\_fXmY}                                                                     \\
{ LOAD\_WITH\_STRIDE}  & { vlse64\_v\_fXmY}                                                            & { \_\_riscv\_vlse64\_v\_fXmY}                                                                    \\
{ FMA\_VF}                 & { vfmacc\_vf\_fXmY}                                                           & { \_\_riscv\_vfmacc\_vf\_fXmY}                                                                   \\
{ FMA\_VV}                & { vfmacc\_vv\_fXmY}                                                           & { \_\_riscv\_vfmacc\_vv\_fXmY}                                                                   \\
{ MUL\_VF}                 & { vfmul\_vf\_fXmY}                                                            & { \_\_riscv\_vfmul\_vf\_fXmY}                                                                    \\
{ MUL\_VV}                & { vfmul\_vv\_fXmY}                                                            & { \_\_riscv\_vfmul\_vv\_fXmY}                                                                    \\
{ STORE}               & { vse64\_v\_fXmY}                                                             & { \_\_riscv\_vse64\_v\_fXmY}                                                                     \\
{ BROADCAST}           & { vfmv\_v\_f\_fXmY}                                                           & { \_\_riscv\_vfmv\_v\_f\_fXmY}                                                                   \\ 
{ REDUCE}              & { vfredusum\_vs\_fXmfY\_fXm1}                                                 & { \_\_riscv\_vfredusum\_vs\_fXmfY\_fXm1}                                                         \\
{ GET\_FIRST}          & { vfmv\_f\_s\_fXm1\_fX}                                                       & { \_\_riscv\_vfmv\_f\_s\_fXm1\_fX}                                                               \\ \hline
\end{tabular}
\end{center}
\end{table*}

\subsection{Data Structures}

\begin{figure*}[t]
\centering
\includegraphics[width=0.7\textwidth]{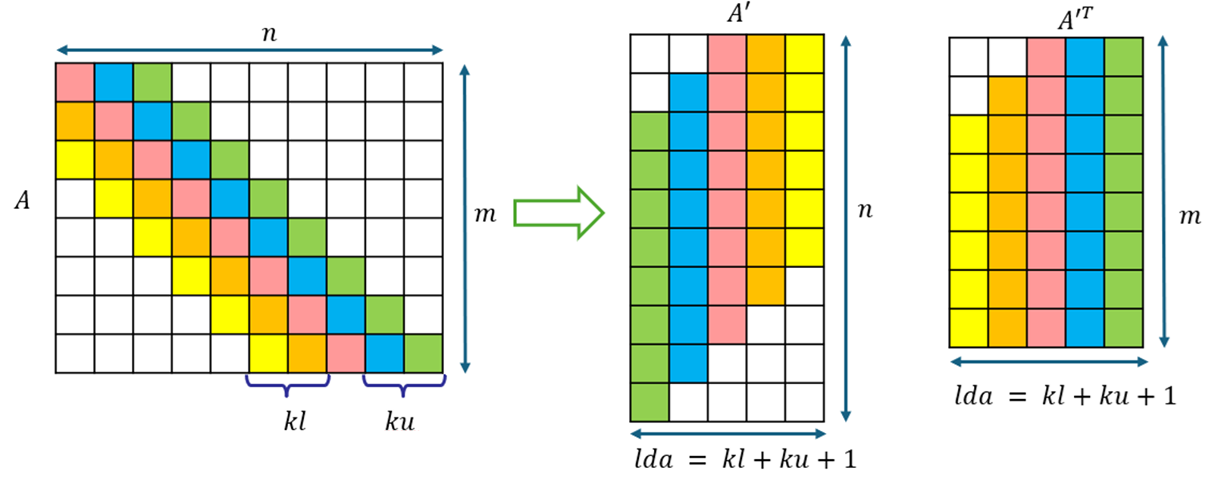}
\caption{Storage format of a general band matrix in the OpenBLAS library. $A$ is a general band matrix, $A'$ and $A'^T$ are dense matrices, which are the storage structures of matrices $A$ and $A^T$, respectively}\label{fig1}
\end{figure*}

\begin{figure*}[t]
\centering
\includegraphics[width=0.7\textwidth]{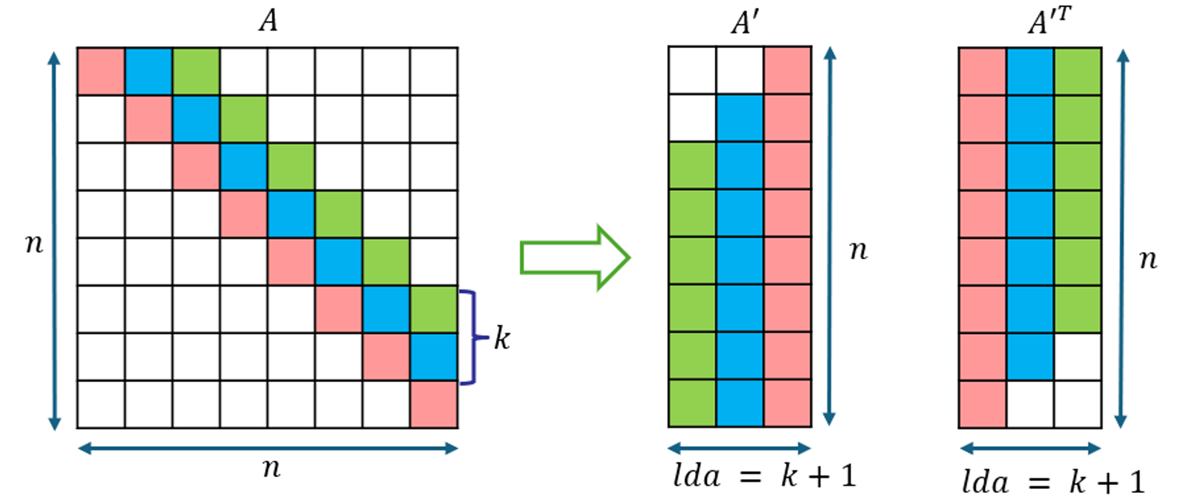}
\caption{Storage format of triangular and symmetric band matrices of general form in the OpenBLAS library. $A$ is an original upper triangular band matrix, $A'$ and $A'^T$ are dense matrices, which are the storage structures of matrices $A$ and $A^T$, respectively}\label{fig2}
\end{figure*}

In the OpenBLAS library, a non-transposed band matrix $A$ of size $m \times n$ with $ku$ upper and $kl$ lower diagonals is stored as a one-dimensional array with indexing corresponding to the column-wise storage of a dense rectangular matrix $A’$ of size $(kl + ku + 1) \times lda$. Here $lda$ is the leading dimension of A. If the matrix is non-transposed, then $lda \geq n$, otherwise $lda \geq m$ (Fig. \ref{fig1}). For a triangular band matrix, only the specified triangle elements are stored (Fig. \ref{fig2}).

Note that the choice of data structure for storing the matrix significantly affects the performance of key algorithms. Thus, to access the column elements of the original matrix $A$, it is necessary to sequentially access the row elements of the matrix $A'$, which is efficient in terms of cache memory usage and convenient for loading into a vector register. The situation is somewhat worse when traversing the diagonal of the original matrix, since in this case it is necessary to load a column from $A’$. This results in memory jumps across $lda$ elements and requires the use of computationally intensive gather instructions (indexed load in RISC-V terminology). A similar problem arises when loading a row of the matrix $A$, which requires working with diagonals of $A’$ in memory, going from the upper right to the lower left corner. These considerations must be taken into account when implementing the underlying algorithms. Typically, these algorithms have low arithmetic intensity and their performance is highly dependent on the efficiency of memory management and the success of vectorization.

\subsection{GBMV: general band matrix-vector multiply}

\subsubsection{Reference algorithm}

Consider the general band matrix-vector multiplication (GBMV) function. According to the BLAS standard, it performs the operation $y = \alpha \times op(A) \times x + \beta y$, where $A$ is a matrix, $op(A)=A$ or $op(A)=A^T$, $x$, $y$ are vectors, and $\alpha$, $\beta$ are scalars. The pseudocode of the baseline version of the GBMV algorithm is shown in Algorithm \ref{alg:1}.

\begin{algorithm}[!ht]
\caption{Pseudocode of the baseline GBMV algorithm in the OpenBLAS library}\label{alg:1}
\begin{algorithmic}[1]

\State void \textbf{GBMV}(INT m, INT n, INT ku, INT kl, FLOAT alpha,
       FLOAT *a, INT lda, FLOAT *X, FLOAT *Y) \{
\State INT offset\_u = ku; INT offset\_l = ku + m;
\For {(i = 0; i < MIN(n, m + ku); i++)}
  \State start   = MAX(offset\_u, 0);
  \State end     = MIN(offset\_l, ku + kl + 1);
  \State length  = end - start;
\EndFor
\If {(not TRANSPOSED)}
\State    \textbf{AXPY}(length, alpha * X[i], a + start, \par
              Y + start - offset\_u); 
\Else
\State    Y[i] += alpha * \textbf{DOT}(length, a + start, \par
              X + start - offset\_u);
\EndIf
\State    offset\_u --; offset\_l --; a += lda;
\State \}
\end{algorithmic}
\end{algorithm}

\begin{figure}[t]
\centering
\includegraphics[width=0.3\textwidth]{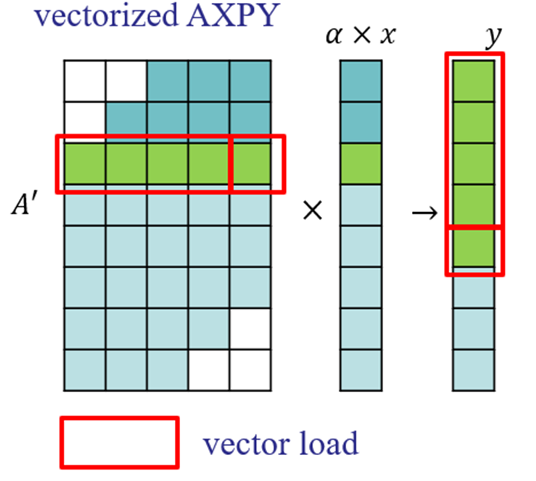}
\caption{Scheme of the baseline algorithm for multiplying a general band matrix by a vector in the OpenBLAS library}\label{fig3}
\end{figure}

The product of $op(A) \times x$ is calculated in a loop over the matrix columns (Algorithm \ref{alg:1}, line 3). For a non-transposed matrix, at each loop iteration, the column of the matrix $A$ is multiplied by the vector $x$ element and the result is added to the corresponding vector $y$ element. For a transposed matrix, the dot product of the matrix $A$ column and the corresponding vector $y$ elements is calculated. In the OpenBLAS library, BLAS level 1 operations are used: AXPY (Algorithm \ref{alg:1}, line 9) and DOT (Algorithm \ref{alg:1}, line 11). These operations are vectorized for various architectures and sets of vector instructions including RVV for RISC-V processors (Fig. \ref{fig3}).

We found that the baseline algorithm performs poorly for narrow-band matrices. When a matrix column has fewer elements than can fit in a vector register, the scalar versions of AXPY and DOT are used, which reduces performance. Otherwise, if the bandwidth is not a multiple of the vector register size, then the inefficiently vectorized tail can be significant for a small bandwidth.

\subsubsection{Optimized algorithm}

\begin{figure}[t]
\centering
\includegraphics[width=0.45\textwidth]{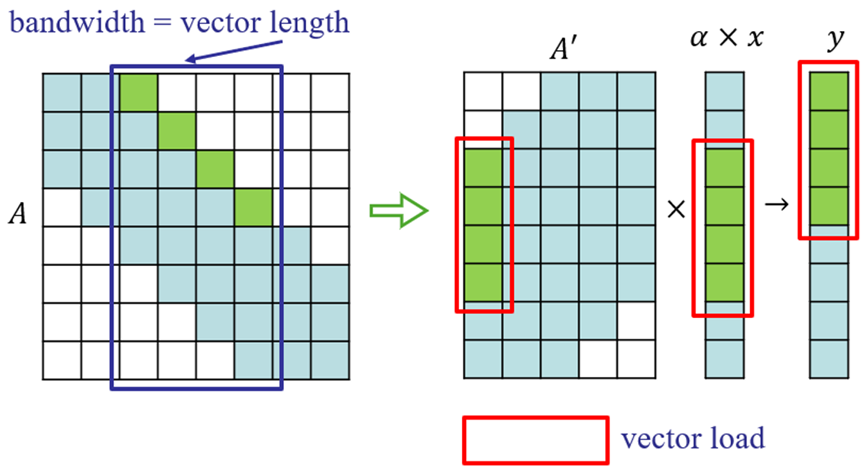}
\caption{Scheme of the optimized algorithm for multiplying a general band matrix by a vector in the OpenBLAS library}\label{fig4}
\end{figure}

For matrices with a narrow bandwidth, a modification of GBMV is proposed that allows for more efficient vectorization by changing the matrix traversal order. To calculate the product $op(A) \times x$ we divide the matrix into vertical blocks, the size of which is equal to the number of elements that fit into the vector register. In each block, a matrix traversal is performed along the diagonals (Fig. \ref{fig4}). Then vector operations of addition and multiplication are performed with the elements of the matrix $A$ diagonal and the corresponding elements of the vectors $x$ and $y$.

The pseudocode for the optimized algorithm is given in Algorithm \ref{alg:2}. This algorithm could be implemented with any set of vector instructions that support addition, multiplication, index load in RISC-V terms, and gather in x86 terms.

\begin{algorithm}[!ht]
\caption{Pseudocode of the optimized GBMV algorithm}\label{alg:2}
\begin{algorithmic}[1]

\State void \textbf{GBMV\_Optimized}(INT m, INT n, INT ku, INT kl, \par 
         FLOAT alpha, FLOAT *a, INT lda,  FLOAT *X, \par FLOAT *Y) \{
\State   INT start, end;
\State   INT BLOCK\_SIZE = GET\_VECTOR\_LENGTH(); 
\State   start = ku; 
\If {(not TRANSPOSED)}
\State      end = MIN(n, m - kl);
\Else
\State      end = MIN(m, n - kl);
\EndIf
\State end -= (end - start) \% BLOCK\_SIZE; 
\State \textbf{/* baseline Algorithm \ref{alg:1} for the columns with indices from 0 to start - 1 */}
\State ptrdiff\_t stride\_x, stride\_y; 
\If {(not TRANSPOSED)}
\State    stride\_x = 0; stride\_y = - ku + j;
\Else
\State    stride\_x = - ku + j; stride\_y = 0;
\EndIf

\For {(i = start; i + BLOCK\_SIZE < end; \par i += BLOCK\_SIZE)} 
\State       x\_copy = LOAD(X + i + stride\_x); 
\State       \correction{x\_copy = MUL\_VF(x\_copy, alpha);}
\For {(j = 0; j < kl + ku + 1; j++)} 
\State       y\_copy = LOAD(Y + i + stride\_y);
\State       diag\_a = LOAD\_WITH\_STRIDE(a + j, STRIDE);
\State       \correction{y\_copy = FMA\_VV(diag\_a, x\_copy, y\_copy);}
\State       STORE(Y + i + stride\_y, y\_copy);
\EndFor 
\State    a += lda * BLOCK\_SIZE;    
\EndFor 

\State \textbf{/* baseline Algorithm \ref{alg:1} for the columns with indices from end to n - 1 */}
\State \}

\end{algorithmic}
\end{algorithm}

In Algorithm \ref{alg:2}, lines 3-10 define the starting and ending indices of the columns of matrix $A$, which will be traversed by the optimized algorithm. Line 11 calls Algorithm \ref{alg:1} to process columns with fewer than $lda$ elements. Lines 13-17 calculate the index shift relative to $i$ for operations with vectors $x$ and $y$. The loop in lines 18-27 performs matrix block multiplication: loading vectors $x$ and $y$ (lines 19, 22), loading a diagonal fragment into a vector register (line 23), calculating the product (line 24), and storing the result in $y$ (line 25). Note that loading a diagonal requires using a vector load instruction with non-unit-stride access since these elements are not consecutive in memory. Line 29 calls Algorithm \ref{alg:1} processing last columns with indices from end to $n - 1$.

\subsection{SBMV: symmetric band matrix-vector multiply}

The approach described above is also used to optimize the symmetric band matrix-vector multiplication (SBMV) and triangular band matrix-vector multiplication (TBMV).

SBMV performs the operation $y = \alpha \times A \times x + \beta \times y$, where $x$, $y$ are vectors, $\alpha$, $\beta$ are scalars, $A$ is a symmetric band matrix of $n \times n$ size with $k$ side diagonals. In the baseline version of SBMV, the product $\alpha \times A \times x$ is calculated for one matrix $A$ column at each iteration, AXPY and DOT are called to update the vector $y$, since only one matrix triangle is stored. Algorithm \ref{alg:3} shows an optimized algorithm for this function in case when the matrix is stored as a lower triangle. In case when it is stored as an upper one, the code differs in terms of shifting indexes in arrays $X$, $Y$, a and the calculating column size formula (lines 5, 20, 25). In lines 4-23 of Algorithm \ref{alg:3}, the first $n - k$ columns of the matrix of size $k$ are traversed. The AXPY and DOT calls are separated, because preliminary performance measurements showed that the DOT takes much longer to execute. In lines 4-7 of Algorithm \ref{alg:3}, the AXPY is executed for the first $n - k$ columns, in lines 8-18 of Algorithm \ref{alg:3} vectorized DOT for the same columns is calculated. The matrix is traversed along the diagonals, similar to the optimized GBMV algorithm. DOT for columns that do not fit into vector registers is performed using the baseline algorithm (lines 19-23). The calculations for the matrix last columns with a smaller size are also performed using the baseline algorithm (lines 24-30).

\begin{algorithm}[!ht]
\caption{Pseudocode of the optimized SBMV algorithm. The matrix is stored as a lower triangle}\label{alg:3}
\begin{algorithmic}[1]

\State void \textbf{SBMV\_L\_Optimized}(INT n, INT k, FLOAT alpha, FLOAT *a,
                        INT lda, FLOAT *X, FLOAT *Y) \{
\State INT BLOCK\_SIZE = GET\_VECTOR\_LENGTH();
\State INT iend = (n - k) - (n - k) \% BLOCK\_SIZE;
\For {(i = 0; i < n - k; i++)}
\State    length = MIN(k, n - i - 1);
\State    AXPY(length + 1, alpha * X[i], a + lda * i, Y + i);
\EndFor
\For {(i = 0; i < iend; i += BLOCK\_SIZE)}  
\State y\_copy =LOAD(y + i); 
\For {(INT j = 0; j < k; j++)}
\State    x\_copy = LOAD(x + i + 1 + j); 
\State    diag\_a = LOAD\_WITH\_STRIDE(a + 1 + j,\par STRIDE); 
\State    \correction{mul = MUL\_VV(x\_copy, diag\_a);} 
\State    \correction{y\_copy = FMA\_VF(y\_copy, alpha, mul);} 
\EndFor
\State    SAVE(y + i, y\_copy); 
\State    a += BLOCK\_SIZE * lda; 
\EndFor
\For {(; i < n - k; i++)} 
\State length = MIN(n - i - 1, k); 
\State Y[i] += alpha * DOT(length, a + 1, X + i  + 1); 
\State a += lda; 
\EndFor
\For {(; i < n; i++)}
\State    length = MIN(n - i - 1, k);
\State    AXPY(length + 1, alpha * X[i], a + k - length,  
\State         Y + i - length);
\State    Y[i] += alpha * DOT(length, a + 1, X + i  + 1);
\State    a += lda;
\EndFor
\State \} 

\end{algorithmic}
\end{algorithm}

\subsection{TBMV: triangular band matrix-vector multiply}

TBMV performs the operation $x=op(A) \times x$, where $x$, $y$ are vectors, $A$ is a triangular band matrix of $n \times n$ size with $k$ side diagonals, and $op(A)=A$ or $op(A)=A^T$. Similar to GBMV, at each iteration the product of the matrix column $op(A)$ and the corresponding vector $x$ element is calculated. The result is then written into the input vector $x$, and the traversal is performed ``bottom-up'' and ``top-down'' for a lower and an upper triangular matrix, respectively. To optimize this function, the same approach was used as for GBMV. The pseudocode for the optimized algorithm for a lower triangular non-transposed matrix is given in Algorithm \ref{alg:4}. It traverses the matrix ``bottom-up''. For the last columns, calculations are performed using the baseline algorithm (lines 4-14), and for the remaining columns, they are performed by an optimized algorithm that traverses along diagonals (lines 15-31).

\begin{algorithm}[!ht]
\caption{Pseudocode of the optimized TBMV algorithm. The matrix is lower triangular and non-transposed}\label{alg:4}
\begin{algorithmic}[1]

\State void \textbf{TBMV\_LN\_Optimized}(INT n, INT k, FLOAT *a,\par                        INT lda, FLOAT *B) \{
\State    INT BLOCK\_SIZE = GET\_VECTOR\_LENGTH();
\State    a += (n - 1) * lda;
\State    INT iend = n - k - (n - k) \% BLOCK\_SIZE;
\For {(i = n - 1; i >= iend; i--)} 
\State       length  = MIN(n - i - 1, k);
\If { (length > 0)}
\State          AXPY(length, B[i], a + 1, 1, B + i + 1);
\EndIf
\If{ (not a unit main diagonal)}
\State          B[i] *= a[0];
\EndIf
\State       a -= lda;
\EndFor
\State    a -= (lda - 1) * BLOCK\_SIZE;
\For {(; i >= 0; i -= BLOCK\_SIZE)}             
\State       length = k;
\State       b\_old = LOAD(B + i);
\If {(not a unit main diagonal)}
\State          diag = LOAD\_WITH\_STRIDE(a, stride\_a);
\State          \correction{z = MUL\_VV(b\_old, diag);}
\State          SAVE(B + i, z);
\EndIf
\For {(j = 1; j < k + 1; j++)}
\State          diag = LOAD\_WITH\_STRIDE (a + j, stride\_a);
\State 	    z = LOAD(B + i + j);
\State 	    \correction{z = FMA\_VV(z, diag, b\_old);}
\State 	    SAVE(B + i + j, z);
\EndFor
\State 	 a -= lda * BLOCK\_SIZE;
\EndFor
\State \}

\end{algorithmic}
\end{algorithm}

\subsection{TBSV: triangular band matrix-vector solve}

\subsubsection{Reference algorithm}

\begin{figure}[t]
\centering
\includegraphics[width=0.45\textwidth]{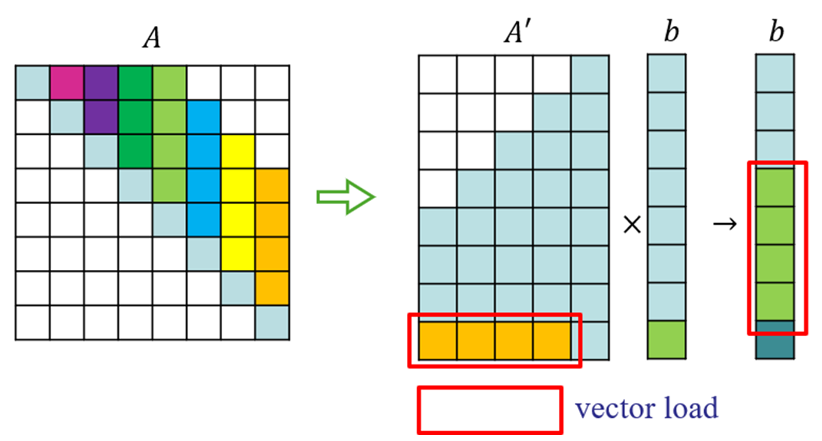}
\caption{Scheme of the solving SLAE algorithm with a TBSV band matrix in the OpenBLAS library.}\label{fig5}
\end{figure}

Consider the function of solving SLAE $op(A) \times x=b$ with a triangular band matrix (TBSV). At each iteration, the scalar product of the matrix $A$ non-diagonal elements and the already found values of the vector of unknows $x$ is subtracted from the right-side element $b[i]$. Then the next unknown $x[i]$ is calculated. (Fig. \ref{fig5}).

Depending on stored triangle of the matrix and whether the matrix is transposed or not, this operation is performed using the DOT or AXPY. A pseudocode of the TBSV algorithm for a lower triangular matrix is given below (Algorithm \ref{alg:5}).

\begin{algorithm}[hbt!]
\caption{Pseudocode of the baseline TBSV algorithm in the OpenBLAS library}\label{alg:5}
\begin{algorithmic}[1]
\State void \textbf{TBSV\_L}(INT n,  INT k, FLOAT *a,  INT lda, \par FLOAT *B) \{
\State    INT i, length;
\For{ (i = 0; i < n; i++)}
\If {(TRANSPOSED)}
\State       length  = MIN(i, k);
\If {(length > 0)} 
\State          B[i] -= \textbf{DOT}(length, a + k - length, 
\State                     B + i - length); 
\EndIf
\State       B[i] /= a[k];
\Else
\State       B[i] /= a[0];
\State       length  = MIN(n - i - 1, k);
\If {(length > 0)} 
\State          \textbf{AXPY}(length, -B[i], a + 1, B + i + 1); 
\EndIf
\State       a += lda;
\EndIf
\EndFor
\State \}
\end{algorithmic}
\end{algorithm}

\subsubsection{Optimized algorithm}

The TBSV optimized algorithm is similar to Algorithm \ref{alg:5} at the top level. The optimization consists of a custom implementation of vector operations, DOT and AXPY. Unlike the baseline implementation, we selected the optimal size of the logical vector registers for these functions (see Section \ref{sec:Methodology} for details). The pseudocode of the optimized TBSV algorithm for lower triangular matrices is shown in Algorithm \ref{alg:6}. Lines 5-15 calculate DOT and lines 24-32 calculate AXPY.

\begin{algorithm}[hbt]
\caption{Pseudocode of the optimized TBSV algorithm for a lower triangular matrix}\label{alg:6}
\begin{algorithmic}[1]

\State void \textbf{TBSV\_L\_Optimized}(INT n, INT k, FLOAT *a, \par INT lda, 
                             FLOAT *B) \{
\If{ (TRANSPOSED)}
\State    \textbf{/* baseline Algorithm \ref{alg:5} for the first k columns */}
\For {(i = k; i < n; i++)}
\State          A\_ptr = a;
\State          B\_ptr = B + i - k;
\State          vsum = BROADCAST(0); 
\For {(int j = k; j > 0; )} 
\State             a\_copy = LOAD(A\_ptr);
\State             b\_copy = LOAD(B\_ptr);
\State             \correction{vsum = FMA\_VV(a\_copy, b\_copy,  vsum);}
\State             step = MIN(j, MAX\_VECTOR\_LENGTH);
\State             A\_ptr += step; B\_ptr += step; j -= step;
\EndFor
\State          \correction{vreduce = REDUCE(vreduce, vsum);}
\State          \correction{dot = GET\_FIRST(vreduce);}
\State          B[i] = (B[i] - dot)/ a[k];
\State          A += lda;
\EndFor
\Else
\For {(i = 0; i <= n - k - 1; i++)}
\State          B[i] /= a[0];
\State          A\_ptr = a + 1;
\State          B\_ptr = B + i + 1;
\For {(j = k; j > 0; )}
\State             a\_copy = LOAD(A\_ptr);
\State             b\_copy = LOAD(B\_ptr);
\State             mult = BROADCAST(-B[i]);
\State             \correction{b\_copy = FMA\_VF(mult, a\_copy, b\_copy);}
\State             STORE(B\_ptr, b\_copy);
\State             step = MIN(j, MAX\_VECTOR\_LENGTH);
\State             A\_ptr += step; B\_ptr += step; j -= step;
\EndFor
\State          a += lda;
\EndFor
\State       \textbf{/* baseline Algorithm \ref{alg:5} for the last k columns */}
\EndIf
\State \}

\end{algorithmic}
\end{algorithm}

\begin{figure*}[ht]
\centering
\includegraphics[width=\textwidth]{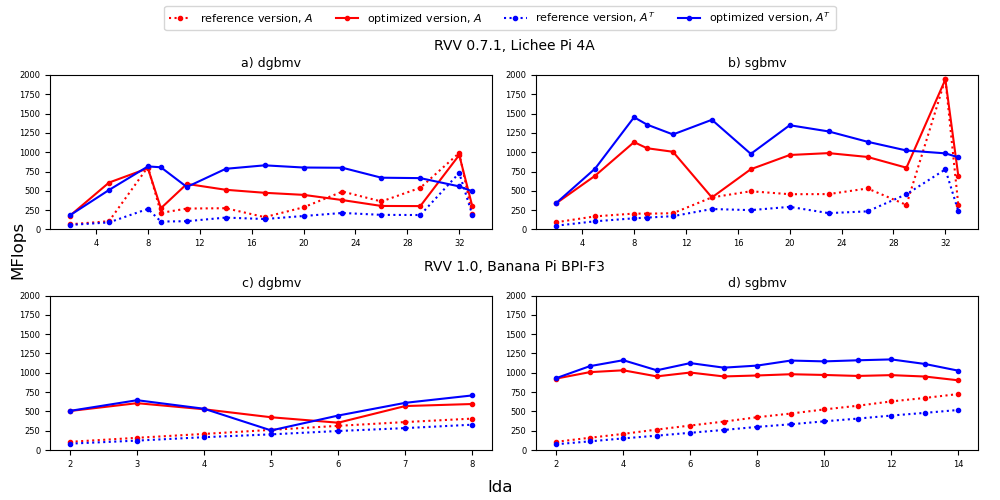}
\caption{Performance of the OpenBLAS reference and our optimized GBMV implementations on 5M-row matrices with different bandwidths: a) double-precision GBMV on Lichee Pi 4A board, b) single-precision GBMV on Lichee Pi 4A board, c) double-precision GBMV on Banana Pi BPI-F3 board, d) single-precision GBMV on Banana Pi BPI-F3 board. All experiments are performed in sequential mode}\label{fig6}
\end{figure*}

\begin{figure*}[ht]
\centering
\includegraphics[width=\textwidth]{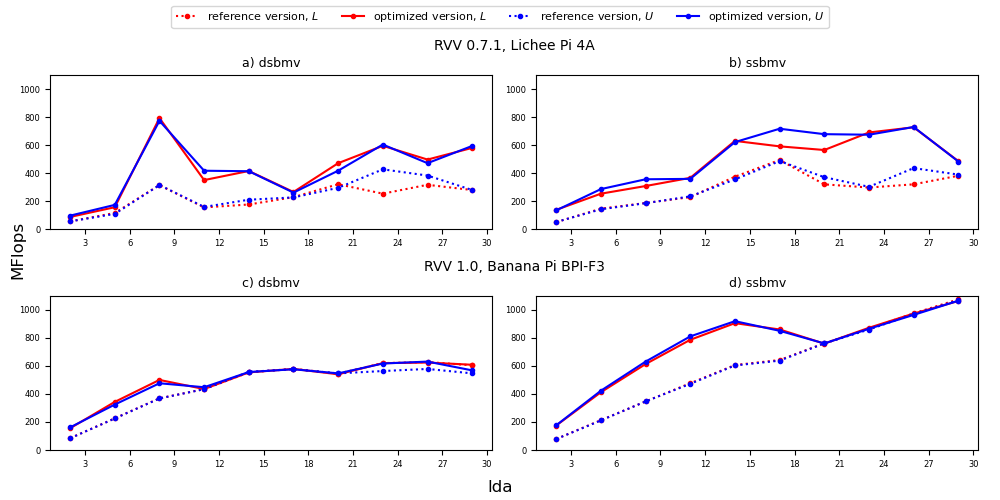}
\caption{Performance of the OpenBLAS reference and our optimized SBMV implementations on 5M-row matrices with different bandwidths: a) double-precision SBMV on Lichee Pi 4A board, b) single-precision SBMV on Lichee Pi 4A board, c) double-precision SBMV on Banana Pi BPI-F3 board, d) single-precision SBMV on Banana Pi BPI-F3 board. All experiments are performed in sequential mode. $L$ -- matrix is stored as lower triangle, $U$ -- matrix is stored as upper triangle}\label{fig7}
\end{figure*}

\section{Numerical results}
\label{sec4}

\subsection{Infrastructure}
Computational experiments were conducted on two types of nodes. The first test system, \textit{Lichee Pi 4A} board, has a T-Head TH1520 processor (4x1.85GHz T-Head C910 cores with 12-stage out-of-order multiple issue superscalar pipeline, ISA RV64GC and 128-bit RVV 0.7.1 standard) with 16 GB of RAM and the Debian GNU/Linux 12 (bookworm) operating system. It uses the gcc (Xuantie-900 linux-5.10.4 Toolchain V2.8.1 B-20240115) 10.4.0 cross compiler. \correction{Lichee Pi 4A theoretical peak double precision -- 14.8 GFLOPs, single precision -- 29.6 GFLOPs, STREAM: DRAM $\sim$8 GB/sec; L2 $\sim$16 GB/sec; L1 $\sim$85 GB/sec.} The second test system, \textit{Banana Pi BPI-F3} board, has a SpacemiT Keystone K1 processor (8x1.6GHz SpacemiT x60 cores with 8-stage in-order dual-issue pipeline, RVA22 Profile and 256-bit RVV 1.0 standard) with 16 GB of RAM and the Bianbu 1.0.15 operating system. It uses the GCC RISC-V 14.2.0 cross-compiler. \correction{Banana Pi  BPI-F3 theoretical peak double precision -- 12.8 GFLOPs, single precision -- 25.6 GFLOPs, STREAM: DRAM $\sim$8 GB/sec; L2 $\sim$8 GB/sec; L1 $\sim$16 GB/sec.}

\subsection {Methodology} \label{sec:Methodology}  
The algorithms described in Section \ref{sec3} were implemented for single and double precision for the AVX-512 vector instruction sets (x86 processors), RVV 0.7.1 and RVV 1.0 vector instruction sets (RISC-V processors), and were integrated into the serial and parallel interfaces of the OpenBLAS library. The optimized code is available at \cite{c67}. Testing was performed using the BLAS-Tester \cite{c61}. For each algorithm, the execution time and performance of the baseline and optimized implementations were compared for matrices ranging from 100,000 to 5 million rows and bandwidth ranging from 1 to 32 for matrix-vector multiplication functions and from 1 to 51 for TBSV. Sequential versions of all functions and parallel versions of matrix-vector multiplications were tested. \correction{For each experiment, the minimum time from the series of runs was selected.} It was found that the algorithm performance does not depend on the matrix size but depends on its bandwidth. Therefore, the plots of the performance dependence on the bandwidth for fixed size matrices are given below. Since RVV allows combining several registers into one logical register and working with the second ones using the $LMUL$ parameter, we empirically selected the optimal value of this parameter for each test system. For matrix-vector multiplication functions, $LMUL=4$ was selected for Lichee Pi 4A and $LMUL=2$ -- for Banana Pi BPI-F3, which corresponds to 512-bit logical registers. TBSV uses $LMUL=1$ for Lichee Pi 4A and $LMUL=2$ for Banana Pi BPI-F3, which corresponds to 128-bit and 512-bit logical registers. Testing parallel versions of matrix-vector multiplication algorithms showed that in multithreaded mode their running time is reduced only for matrices of about 5 million rows and a bandwidth of at least 20. Since our implementations are optimized for matrices with a narrow bandwidth, using parallelization here is not practical. Additionally, sequential implementations of BLAS functions are in demand when solving many problems using numerical modeling methods, where parallel computations at a higher algorithm level are used. In any case, effective vectorization of basic mathematical functions is a necessary condition for utilization of the processor resources.

For brevity, we denote the versions of the functions intended for the lower triangular non-transposed matrices as $LN$, the lower triangular transposed -- $LT$, the upper triangular non-transposed -- $UN$, the upper triangular transposed -- $UT$. In the plots, the bandwidth is denoted by $lda$.

\subsection{GBMV Algorithm}
At first, we consider GBMV. The results of computational experiments for the sequential version of the function on matrices with 5 million rows and different bandwidth are shown in Figure \ref{fig6}. GBMV was implemented in two versions, for the non-transposed and for the transposed matrices.

On Lichee Pi 4A (RVV 0.7.1), the optimized algorithm for double-precision matrices outperforms the baseline for non- transposed matrices with fewer than 20 diagonals and for transposed ones with any bandwidth. The average speedup is 2.4x and 4.2x, respectively. For single-precision matrices, the optimized version significantly outperforms the baseline for matrices of any size, showing a performance 2.9x and 5.6x greater for non-transposed and transposed matrices, respectively.

For Banana Pi BPI-F3 (RVV 1.0), the optimized version outperforms the baseline when the bandwidth is less than or equal to 8 for double-precision and less than or equals to 14 in single-precision matrices. The average speedup is 2.4x and 3.2x for double-precision and 3.1x and 4.9x for single-precision non-transposed and transposed matrices, respectively.

In general, the optimized version outperforms the baseline one for matrices with less than 14 diagonals in most test configurations on both systems. In all test configurations, a great relative speedup (up to 10x) is obtained on transposed matrices, where DOT was called in the baseline algorithm. The average speedup is slightly greater for single-precision matrices than for double-precision ones.

\subsection{SBMV Algorithm}

The results of computational experiments for the SBMV sequential version are shown in Figure \ref{fig7}. We consider matrices with 5 million rows and different bandwidth. SBMV is implemented in two versions, for transposed and non-transposed matrices.

The plots show that the speedup for non-transposed matrices does not exceed 2.5x and averages 1.8x and 1.3x for Lichee Pi 4A (RVV 0.7.1) and Banana Pi BPI-F3 (RVV 1.0) test systems, respectively. On Lichee Pi 4A the optimized version outperforms the baseline one for any $lda$ value. For Banana Pi BPI-F3, it outperforms up to $lda=14$ for double-precision and up to $lda=20$ for single-precision matrices. For a larger bandwidth, the optimized version slows down compared to the baseline one. Therefore, for RVV 1.0, switching to the reference version depending on the bandwidth was added. In general, for RVV 0.7.1, the optimized version works stably in all test configurations. For RVV 1.0, it works stably for matrices with a bandwidth of less than 14, but the average performance speedup is less than that of the GBMV function.

\subsection{TBMV Algorithm}

\begin{figure*}[ht]
\centering
\includegraphics[width=\textwidth]{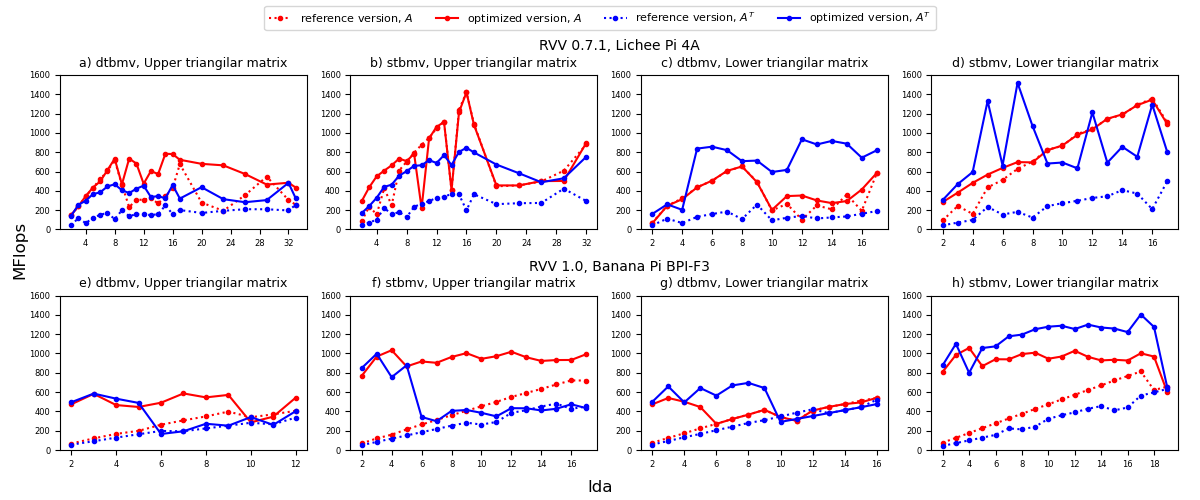}
\caption{Performance of the OpenBLAS reference and our optimized TBMV implementations on 1M-row matrices with different bandwidths. The results on Lichee Pi 4A board are shown in plots a)-d), the results on Banana Pi BPI-F3 board are shown in plots e)-h). All experiments are performed in sequential mode.
a) double-precision TBMV for upper triangular matrices (UN and UT case), b) single-precision TBMV for upper triangular matrices (UN and UT case), c) double-precision TBMV for lower triangular matrices (LN and LT case), d) single-precision TBMV for lower triangular matrices (LN and LT case), e) double-precision TBMV for upper triangular matrices (UN and UT case), f) single-precision TBMV for upper triangular matrices (UN and UT case), g) double-precision TBMV for lower triangular matrices (LN and LT case), h) single-precision TBMV for lower triangular matrices (LN and LT case)
}\label{fig8}
\end{figure*}

Here we consider TBMV. The results of computational experiments for the TBMV sequential version are shown in Figure \ref{fig8}. The plots were constructed for matrices with 1 million rows and different bandwidths.

This function is implemented in four versions: for upper and lower triangles, transposed and non-transposed matrices (LN, LT, UN, UT). The optimized version replaces DOT in the baseline algorithm in versions UT and LT, and provides the greatest performance speedup. For TBMV on Lichee Pi 4A (RVV 0.7.1) these versions outperform the baseline ones for any bandwidth and provide a greater advantage for the smaller bandwidth. The average advantage is 2.5x with the UT, 5.2x with the LT version for double-precision matrices and 2.6x and 4.5x for single-precision ones, respectively. On Banana Pi BPI-F3 (RVV 1.0) the performance of optimized UT and LT is higher than the baseline one with any bandwidth for single-precision matrices and with narrow bandwidth in double-precision ones. In almost all test cases for bandwidth of 2 and 3, the optimized algorithm significantly outperforms the baseline, by 2x to 13x on RVV 0.7.1 and by 4x to 20x on RVV 1.0. However, the least effect of the optimization is seen for the LN version, where the optimized algorithm replaced AXPY and traversed the matrix ``bottom-up''. On Banana Pi BPI-F3, the optimized algorithm outperforms the baseline one only for matrices with no more than 6 diagonals in single and double precision. On Lichee Pi 4A, LN optimized version works better on narrow bandwidth single-precision matrices, for double-precision ones -- when the bandwidth is greater than 8. For UN versions on Lichee Pi 4A, the optimized algorithm outperforms baseline one for wide double-precision matrices (when \correction{bandwidth} is greater than 8) and narrow (\correction{bandwidth} is less than 5) or wide (\correction{bandwidth} is 20 and more) single-precision matrices. On Banana Pi BPI-F3, the optimized version outperforms the baseline by an average of 2x and 2.8x for any bandwidth double- and single-precision matrices, respectively.

\subsection{TBSV Algorithm}

\begin{figure*}[ht]
\centering
\includegraphics[width=\textwidth]{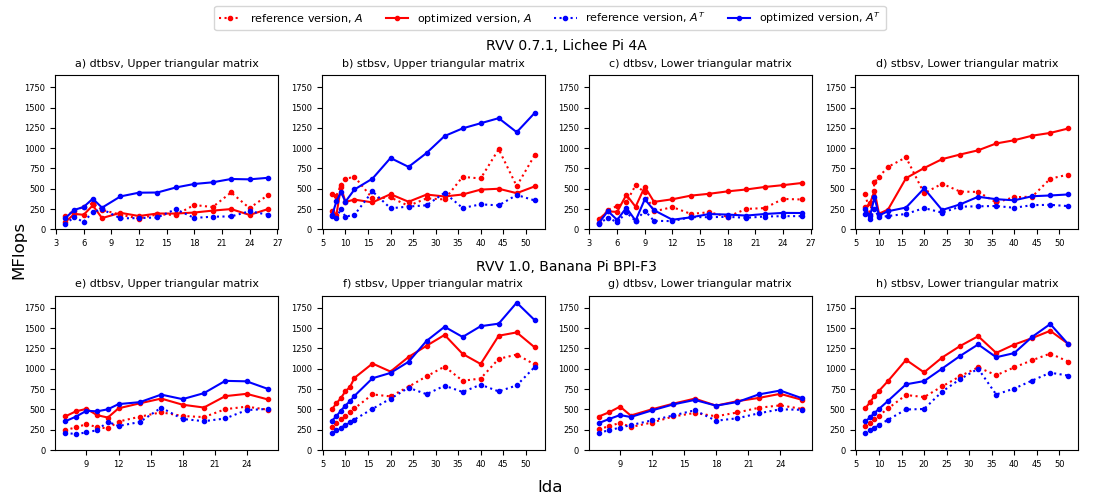}
\caption{Performance of the OpenBLAS reference and our optimized TBSV implementations on 250 000-row matrices with different bandwidths. The results on Lichee Pi 4A board are shown in plots a)-d), the results on Banana Pi BPI-F3 board are shown in plots e)-h). All experiments are performed in sequential mode.
a) double-precision TBSV for upper triangular matrices (UN and UT case), b) single-precision TBSV for upper triangular matrices (UN and UT case), c) double-precision TBSV for lower triangular matrices (LN and LT case), d) single-precision TBSV for lower triangular matrices (LN and LT case), e) double-precision TBSV for upper triangular matrices (UN and UT case), f) single-precision TBSV for upper triangular matrices (UN and UT case), g) double-precision TBSV for lower triangular matrices (LN and LT case), h) single-precision TBSV for lower triangular matrices (LN and LT case)
}\label{fig9}
\end{figure*}

The results of the computational experiments for TBSV are shown in Figure \ref{fig9}. The plots are given for matrices with 250000 rows and different bandwidths.

TBSV is implemented in four variants: for the upper and lower triangles, transposed and non-transposed matrices (LN, LT, UN, UT).

The plots show that the optimized implementation on both test boards outperforms the baseline version for both variants (UT and LT) where the baseline implementation calls DOT. On the Lichee Pi 4A (RVV 0.7.1) board, the average performance speedup of the UT version is 2.8x for double- and 3.0x for single-precision matrices. For the LT version, it is 1.3x faster for both single and double precision. On the Banana Pi BPI-F3 (RVV 1.0) board, the optimized version is 1.5x and 1.8x faster than the baseline one for UT and LT, respectively, with both single and double precision. On this board, LN and UN optimized versions outperform the baseline implementation for all test cases and work on average 1.5x faster. The optimization also improves performance on the Lichee Pi 4a board for these TBSV variants. The optimized LN version outperforms the baseline one for double-precision and single-precision matrices with a bandwidth of at least 9 and 20, respectively. It wins on average by 1.8x in double and 2.2x in single precision compared to the baseline version. The UN version performs better on matrices of average bandwidth. It outperforms the baseline by 13\% in both single and double precision.

We conclude that vectorization performed on RVV 1.0 speeds up all TBSV variants. Vectorization on RVV 0.7.1 allows for speeding up the baseline version for transposed matrices with any bandwidth and non-transposed ones with a certain bandwidth. The speedup is less than for other functions, since the algorithmic capabilities of vectorization are limited here.

\section{Conclusion}
\label{sec5}

In this paper, we optimized four BLAS level 2 functions for RISC-V processors: matrix-vector multiplication for general, triangular, and symmetric band matrices, and solving SLAE with a triangular band matrix. Compared to the OpenBLAS RISC-V baseline implementation, our optimization achieves an advantage by changing the matrix traversal order from element-wise to diagonal, which increases vectorization efficiency. The implementation is based on OpenBLAS and is intended for RISC-V processors that support the vector extensions RVV 0.7.1 and RVV 1.0.

Overall, the results of experiments show that, for all four functions, the optimized implementation performance is better than the baseline for matrices with a certain bandwidths or in all cases considered. For all functions, the optimized version provides the greatest speedup for upper triangular or transposed matrices in cases where DOT is called from the baseline algorithm. The switching thresholds between the base and optimized implementations can be determined empirically depending on the input matrix bandwidth. \correction{Note that our implementation mainly demonstrates an advantage for band matrices with a sufficiently small number of diagonals. The gain is achieved mainly due to vectorization, while the basic implementation from OpenBLAS does not cope with efficient vectorization due to the use of a suboptimal data traversal. It is important to note that in most cases we make a better choice between baseline and optimized implementations depending on the algorithm parameters and the number of matrix diagonals. Therefore, the user can run the algorithms as a black box, without delving into the process of choosing the most efficient implementation. However, if there is a significant change in future hardware parameters, we recommend experimenting and adjusting the thresholds if necessary.} It is \correction{also} advisable to use the parallel version of OpenBLAS for matrices with a number of rows of at least 5 million and a large bandwidth (more than 20). In other cases, it is more profitable to use parallelism at a higher level.

The developed experimental implementation is publicly available in \cite{c67}. In the future, we plan to integrate it into the OpenBLAS library and continue research on improving the efficiency of cache memory usage, for example, using the approaches proposed in \cite{c43}. Another area of interest is the development of analytical models for automatic selection of algorithm parameters, such as switching thresholds between different implementations, which can be implemented using machine learning algorithms. As specialized instruction sets for sparse matrix operations become available in the RISC-V architecture, a promising area of research is emerging that is important for practical use in Computational Science. Overall, the prospects for using RISC-V processors for high-performance computing seem undeniable and open up a large field for new research and development both in the area of system programming (development of compilers and profilers) and in the area of porting and optimizing specific software frameworks.

\section*{Acknowledgements}
The project is supported by the Lobachevsky University academic excellence program ``Priority-2030''. The authors acknowledge the use of computational resources provided by the University (Lobachevsky Supercomputer).


 \bibliographystyle{elsarticle-harv} 
 \bibliography{cas-refs}






\end{document}